# Memristive Systems Analysis of 3-Terminal Devices

Blaise Mouttet
George Mason University
Fairfax, Va USA

*Abstract*— **Memristive systems were proposed in 1976 by Leon Chua and Sung Mo Kang as a model for 2-terminal passive nonlinear dynamical systems which exhibit memory effects. Such systems were originally shown to be relevant to the modeling of action potentials in neurons in regards to the Hodgkin-Huxley model and, more recently, to the modeling of thin film materials such as $TiO_{2-x}$ proposed for non-volatile resistive memory. However, over the past 50 years a variety of 3-terminal non-passive dynamical devices have also been shown to exhibit memory effects similar to that predicted by the memristive system model. This article extends the original memristive systems framework to incorporate 3-terminal, non-passive devices and explains the applicability of such dynamic systems models to 1) the Widrow-Hoff memistor, 2) floating gate memory cells, and 3) nano-ionic FETs.**

## I. INTRODUCTION

Memristive systems [1] was originally defined in 1976 by Leon Chua and Sung Mo Kang as an extension of the memristor [2]. The main feature which distinguished memristive systems from the broader class of dynamical non-linear systems is a zero-crossing property in which the output of the system is zero whenever the input is zero. Such systems were defined mathematically as:

$$\frac{dw}{dt} = f(w,u,t)$$
$$y = g(w,u,t)u \qquad (1)$$

where $t$ denotes time, $u$ and $y$ denote the input and output of the system, and $w$ is an n-dimensional vector representing the state of the system. The function $f$ was defined as a continuous n-dimensional vector function while $g$ was defined as a continuous scalar function.

The original memristive systems paper proved several generic properties related to the passivity, frequency response, and small signal AC characteristics of devices or systems which meet the form of (1). It was also shown that existing devices and systems such as the thermistor, discharge tubes, and the Hodgkin-Huxley circuit model of nerve axon membranes may be modeled using the memristive framework. Despite the formal development of memristive systems it has rarely been applied in electronic device design or analysis. Only recently has it been recognized that the memristive systems framework may be relevant to the modeling of passive, 2-terminal semiconductor devices which exhibit memory resistance effects [3]. This is despite the fact that materials exhibiting such effects were known [4] several years prior to the original memristive systems paper !

The memristive systems framework has recently been expanded to cover memory capacitance and memory inductance devices [5]. This begs the question of whether it would also be useful to develop an expanded memristive systems model to cover memory transistors. Several examples of memory transistors are found in the literature dating back 50 years [6]-[16] but there is an apparent lack of a formal dynamic systems analysis of these devices comparable to the memristive systems approach of Chua and Kang. In Section II formal dynamic systems framework for memory transistors is proposed and small-signal AC characteristics are determined. In Section III some examples of memory transistors from the literature are discussed in terms of this framework.

## II. MEM-TRANSISTORS: DEFINITIONS AND SMALL SIGNAL ANALYSIS

A voltage-controlled n-th order mem-transistor is defined by the systems equations:

$$\frac{dw}{dt} = f(w, v_g, v_d)$$
$$i_g = g(w, v_g, v_d) \qquad (2)$$
$$i_d = h(w, v_g, v_d)$$

where $v_g$ and $v_d$ denote input voltages, $i_g$ and $i_d$ denote output currents, and $w$ is an n-dimensional vector representing the state of the system. The function $f$ is defined as a continuous n-dimensional vector function while $g$ and $h$ are defined as continuous scalar functions.

The system expressed by (2) is a 2-port dynamical system. The case of multiport memristive systems was contemplated in [1] but not analyzed. The more significant distinction between (2) and the general equations of a memristive system is the lack of the zero-crossing property. This more general approach eliminates the passivity requirement found in memristive systems but allows for the possibility of modeling active devices capable of both signal amplification and memory storage.

In the same manner as (2) a current-controlled mem-transistor is defined by the systems equations:

$$\frac{dw}{dt} = f(w, i_g, i_d)$$
$$v_g = g(w, i_g, i_d) \quad (3)$$
$$v_d = h(w, i_g, i_d)$$

wherein the currents are the inputs and the voltages are the outputs. Similarly, a hybrid voltage-current controlled mem-transistor is defined as:

$$\frac{dw}{dt} = f(w, v_g, i_g)$$
$$v_d = g(w, v_g, i_g) \quad (4)$$
$$i_d = h(w, v_g, i_g).$$

The small signal AC characteristics of mem-transistors can be determined by the linearization method in a similar manner to that discussed in Property 7 of [1]. One may take as a representative example a 1$^{st}$ order voltage-controlled mem-transistor as in (2) in which $v_d$ is a fixed DC voltage. Linearization about state $W_0$ and voltage $V_{g0}$ produces:

$$\delta\left(\frac{dw}{dt}\right) = \frac{\partial f(W_0, V_{g0})}{\partial w}\delta w + \frac{\partial f(W_0, V_{g0})}{\partial v_g}\delta v_g$$
$$\delta(i_g) = \frac{\partial g(W_0, V_{g0})}{\partial w}\delta w + \frac{\partial g(W_0, V_{g0})}{\partial v_g}\delta v_g \quad (5)$$
$$\delta(i_d) = \frac{\partial h(W_0, V_{g0})}{\partial w}\delta w + \frac{\partial h(W_0, V_{g0})}{\partial v_g}\delta v_g.$$

Applying the Laplace Transform to (5) one obtains:

$$s\Delta W(s) = \frac{\partial f(W_0, V_{g0})}{\partial w}\Delta W(s) + \frac{\partial f(W_0, V_{g0})}{\partial v_g}\Delta V_g(s)$$
$$\Delta I_g(s) = \frac{\partial g(W_0, V_{g0})}{\partial w}\Delta W(s) + \frac{\partial g(W_0, V_{g0})}{\partial v_g}\Delta V_g(s) \quad (6)$$
$$\Delta I_d(s) = \frac{\partial h(W_0, V_{g0})}{\partial w}\Delta W(s) + \frac{\partial h(W_0, V_{g0})}{\partial v_g}\Delta V_g(s)$$

The transconductance $g_m$ of the 1$^{st}$ order voltage-controlled mem-transistor can be determined by solving the first equation for the state $\Delta W(s)$ and plugging the result into the equation for $\Delta I_d(s)$ obtaining:

$$g_m = \frac{\Delta I_d(s)}{\Delta V_g(s)} = \frac{\frac{\partial h(W_0, V_{g0})}{\partial w}\frac{\partial f(W_0, V_{g0})}{\partial v_g}}{s - \frac{\partial f(W_0, V_{g0})}{\partial w}} + \frac{\partial h(W_0, V_{g0})}{\partial v_g}. \quad (7)$$

As a result of (7) some general properties of mem-transistors can be stated:

- For periodic excitation frequencies $f$ ($s=j2\pi f$) the transconductance of a mem-transistor is generally a frequency dependent complex number and represents both gain *and a phase shift* between the input and output signals.

- At high excitation frequencies (f→∞) the first term of (7) reduces to zero and the transconductance reduces to that of an ordinary transistor. This is analogous to the limiting resistive characteristic of memristive systems at high frequency, i.e. property 6 of [1].

- Stability of the mem-transistor requires

$$\frac{\partial f(W_0, V_{g0})}{\partial w} \leq 0 \quad (8)$$

since otherwise the inverse Laplace transform of the impulse response would produce a growing exponential.

III. MEM-TRANSISTOR EXAMPLES

*A. Widrow-Hoff Memistor*

In 1960 Bernard Widrow of Stanford University was working on developing adaptive circuitry having the capability to simulate neurons [6]. His graduate student Marcian Hoff suggested using electroplating cells as a means of developing an electrically controlled variable resistance device. The term "memistor" (i.e memory resistor) was used in describing the 3-terminal cell. As described by Widrow in [6]:

*An ideal memistor would have the following electrical characteristics: the conductance would vary linearly with total plating charge. Achieving this characteristic requires that the plating process be reversible, that the memistor resistance stay put indefinitely when plating current is zero, that the conductance vary smoothly with plating current, and that there be no hysteresis associate with change in direction of plating.*

Widrow was able to achieve very near to this ideal memistor using a pencil lead in a copper sulfate-sulfuric acid plating bath having brightener additives to eliminate unwanted hysteresis. Based upon Fig. 6b of [6] a model for Widrow's memistor may be formed in the linear region of the conductance vs. charge curve. In this region the conductance of the memistor may be approximated as the ratio of the current

flowing through the memistor $i_d$ and the voltage across the memistor $v_d$ so that:

$$\frac{i_d}{v_d} \approx (10^{-4} mhos/C) q_g \qquad (9)$$

wherein $q_g$ is the plating charge. Equivalently to (9) mem-transistor equations may be written in terms of the plating current $i_g$ with the plating charge equated to the state variable $w$.

$$\frac{dw}{dt} = i_g$$
$$v_d \approx \frac{i_d}{(10^{-4} mhos/C) w} \qquad (10)$$

A similar linearization procedure as in Section II may be performed for a memistor having a fixed dc current $i_d = I_{DC}$ biased around a plating charge $Q_{g0}$ and plating current $I_{g0}$. In this case the small signal transimpedance may be calculated as:

$$\frac{\Delta V_d(s)}{\Delta I_g(s)} \approx -\frac{I_{DC}}{(10^{-4} mhos/C) Q_{g0}^2 s}. \qquad (11)$$

It is notable that the form of transfer function (11) is familiar from control theory as that of an ideal integrator.

### B. Floating gate synaptic transistor

A floating gate MOSFET was developed for analog learning applications as described by Diorio, Hasler, Minch, and Mead [10]. Their paper derived dynamic state equations (12) based on the sub-threshold operation of the transistor and in consideration of hot electron injection and Fowler-Nordheim tunneling. These equations may be understood as falling into the class of a 1st order voltage controlled mem-transistor with the gate-to-channel potential $V_{gc}$ as the control voltage and the other parameters considered as constants determined by operating temperature, bias condition, or transistor fabrication details such as doping level and channel dimensions. In these equations the state variable $w$ is equated with the source current.

$$\frac{dw}{dt} \approx \frac{\Delta w_{max} w^{1-\sigma}}{w_{corner} + w^{1-\sigma}} - \frac{\kappa \eta}{Q_T} w^2 e^{-\frac{V\alpha}{V_{gc}}\left(\frac{V\beta}{V_{dc}+V\eta}\right)^2}$$

$$i_g = \varepsilon(V_{dg} + V_{bi})^2 e^{-\frac{V0}{V_{dg}+V_{bi}}} - \eta w e^{-\frac{V\alpha}{V_{gc}}\left(\frac{V\beta}{V_{dc}+V\eta}\right)^2} \qquad (12)$$

$$i_d = w\left(1 + \gamma e^{-\sqrt{\frac{V_m}{V_{dc}-V\gamma}}}\right)$$

It is straightforward (although cumbersome) to calculate the transconductance of this mem-transistor using (7). Applying the small-signal stability criteria (8) to the dynamic equation of (12) linearized around $w_0$ and $V_{gc0}$ produces an inequality in which the state variable terms can be isolated on one side so that:

$$\frac{w_{corner} \Delta w_{max} w_0^{-1-\sigma}(1-\sigma)}{\left(w_{corner} + w_0^{1-\sigma}\right)^2} \leq \frac{2\kappa\eta}{Q_T} e^{-\frac{V\alpha}{V_{gc0}}\left(\frac{V\beta}{V_{dc}+V\eta}\right)^2} \qquad (13)$$

As described in [10] $\sigma$ is linearly related to the thermal voltage and thus increases with temperature. Rearranging (13) to consolidate all constant terms into a single term $K$ and emphasizing the dependence of $\sigma$ on temperature produces:

$$w_0^{1+\sigma(T)}\left(w_{corner} + w_0^{1-\sigma(T)}\right)^2 \geq K(1-\sigma(T)) \qquad (14)$$

At low temperatures when $\sigma(T)<1$ the above equation indicates that the state variable $w$ (i.e. the source current) needs to be larger than a critical value for stability of the transistor. However, at high temperatures when $\sigma(T)>1$ the source current needs to be smaller than a critical value to maintain transistor stability.

### C. Nanoionic FETs

Memristive systems models have been formulated for thin film metal oxides based on an ionic drift mechanism in which oxygen vacancies act as positive ions and the rate of change of the state variable is equated with the ionic drift rate [3]. Similar nanoionic thin films have been constructed based on chalcogenide and other materials [17]. Although nanoionic films have been described in relation to 2-terminal devices it may also be applicable to transistor designs similar to a MOSFET but in which the gate oxide is replaced in part by the nanoionic film. For example, one may contemplate a dual layer gate dielectric film with one of the layers having mobile ions and acting as a conductor and the other layer excluding the mobile ions and acting as a dielectric. Transistors that appear to fall into this category are described in [7] and [16]. This type of gate dielectric structure may be modeled as a variable capacitor with the motion of the ions changing the effective dimensions of the capacitor dielectric region. Using $w$ to represent the effective width of the dielectric region mem-transistor equations (15) may be written for this type of FET operating in the linear (triode) region and assuming no gate current leakage.

$$\frac{dw}{dt} \approx fae^{-U/kT} \sinh\left(\frac{qE(V_{GC},w)a}{2kT}\right)$$

$$i_g \approx 0 \qquad (15)$$

$$i_d = \mu_n C(w) \frac{W}{L}\left[(V_{GS} - V_T(w))V_{DS} - 0.5V_{DS}^2\right]$$

The dynamic equation for $dw/dt$ is based on the nonlinear ionic drift equation of [18] in which the electric field $E$ is a function of both the gate-to-channel voltage $V_{GC}$ and the effective width $w$ of the dielectric region. For example, in the ideal case where the width $w$ is uniform and the nonionic layer is a perfect conductor:

$$E(V_{GC}, w) = \frac{V_{GC}}{w}. \quad (16)$$

The equation for the drain current $i_d$ in (15) is typical for a MOSFET operating in the linear region except that the gate oxide capacitance $C$ and the threshold voltage $V_T$ are no longer constants but are dependent on the ionic distribution. In the ideal case of uniform width w and a perfectly conductive nanoionic layer:

$$C(w) = \frac{\varepsilon_d}{w}. \quad (17)$$

$$V_T(w) = V_{FB} + 2|\phi_p| + \frac{Q_d}{C(w)} \quad (18)$$

wherein $V_{FB}$ is the flatband voltage of the transistor, $\phi_p$ is the bulk potential (for p-type silicon), and $Q_d$ is the depletion charge density. It is straightforward to determine the transconductance of this mem-transistor using (7). The stability criteria (8) produces:

$$-fae^{-U/kT}\cosh\left(\frac{qV_{GC}a}{2kTw}\right)\frac{qV_{GC}a}{2kTw^2} \leq 0 \quad (19)$$

Since this inequality is valid for all w it is indicative that nanoionic FETs may have advantageous stability in comparison to the synaptic floating gate transistors.

## IV. CONCLUSION

The concept of mem-transistor systems has been presented applying some of the techniques from the memristive systems approach of Chua and Kang to non-passive, 2-port dynamic circuit elements. It was shown that a small signal analysis of a voltage-controlled mem-transistor produces an equation for transconductance that is represented as a complex number under periodic excitation and can be used to provide an indication of mem-transistor stability. It was shown that the Widrow-Hoff memistor, the floating gate synaptic transistor, and nanoionic FETs may be modeled using mem-transistor equations and the results of such modeling can be used to determine the small-signal transfer function and perform stability analysis for these devices.